\documentstyle[11pt]{article}
\catcode`\@=11
\@addtoreset{equation}{section}
\def\theequation{\arabic{section}.\arabic{equation}}
\def\appendix{\renewcommand{\thesection}{\Alph{section}}\setcounter{section}{0}
              \renewcommand{\theequation}
            {\mbox{\Alph{section}.\arabic{equation}}}\setcounter{equation}{0}}
\def\maketitle{\thispagestyle{empty}\setcounter{page}0\newpage
                \renewcommand{\thefootnote}{\arabic{footnote}}
                  \setcounter{footnote}0}
\renewcommand{\thanks}[1]{\renewcommand{\thefootnote}{\fnsymbol{footnote}}
               \footnote{#1}\renewcommand{\thefootnote}{\arabic{footnote}}}
\newcommand{\preprint}[1]{\hfill{\sl preprint - #1}\par\bigskip\par\rm}
\renewcommand{\title}[1]{\begin{center}\Large\bf #1\end{center}\rm\par\bigskip}
\renewcommand{\author}[1]{\begin{center}\Large #1\end{center}}
\newcommand{\address}[1]{\begin{center}\large #1\end{center}}

\def\dinfn{\smallskip Dipartimento di Fisica, Universit\`a di Trento\\
                           and Istituto Nazionale di Fisica Nucleare,\\
                                   Gruppo Collegato di Trento, Italia}

\def\Idinfn{\address{\dinfn}}

\newcommand{\email}[1]{e-mail: \sl #1@science.unitn.it\rm}

\newcommand{\femail}[1]{\thanks{\email{#1}}}
\newcommand{\pacs}[1]{\smallskip\noindent{\sl PACS numbers:
                       \hspace{0.3cm}#1}\par\bigskip\rm}
\def\babs{\hrule\par\begin{description}\item{Abstract: }\it}
\def\eabs{\par\end{description}\hrule\par\medskip\rm}
\renewcommand{\date}[1]{\par\bigskip\par\sl\hfill #1\par\medskip\par\rm}
\newcommand{\ack}[1]{\par\section*{Acknowledgments} #1}
\newcommand{\s}[1]{\section{#1}}
\renewcommand{\ss}[1]{\subsection{#1}}

\newcommand{\ca}[1]{{\cal #1}}         
\def\hs{\qquad}               
\def\nn{\nonumber}            
\def\beq{\begin{eqnarray}}    
\def\eeq{\end{eqnarray}}      
\def\ap{\left.}               
\def\at{\left(}               
\def\aq{\left[}               
\def\ag{\left\{}              
\def\cp{\right.}              
\def\ct{\right)}              
\def\cq{\right]}              
\def\cg{\right\}}             
\def\R{{\hbox{{\rm I}\kern-.2em\hbox{\rm R}}}}   
\def\H{{\hbox{{\rm I}\kern-.2em\hbox{\rm H}}}}   
\def\N{{\hbox{{\rm I}\kern-.2em\hbox{\rm N}}}}   
\def\C{{\ \hbox{{\rm I}\kern-.6em\hbox{\bf C}}}} 
\def\Z{{\hbox{{\rm Z}\kern-.4em\hbox{\rm Z}}}}   
\def\ii{\infty}                                  
\newcommand{\fr}[2]{\mbox{$\frac{#1}{#2}$}}      
\def\tr{\mathop{\rm tr}\nolimits}                  
\def\Tr{\mathop{\rm Tr}\nolimits}                  
\renewcommand{\Re}{\mathop{\rm Re}\nolimits}       
\def\lap{\Delta}                                   

\def\be{\beta}
\def\ga{\gamma}

\def\ep{\varepsilon}
\def\ze{\zeta}

\def\la{\lambda}

\def\si{\sigma}

\def\Ga{\Gamma}

\def\La{\Lambda}

\begin{document}

\preprint{UTF 406}

\title{Quantum Correction to the Entropy of the
      (2+1)-Dimensional Black Hole}

\author{Andrei A. Bytsenko\thanks{email: abyts@fisica.uel.br\,\,\,\,\,
On leave from St. Petersburg State Technical University, Russia}}
\address{Departamento de Fisica, Universidade Estadual de Londrina, Caixa 
Postal 6001, Londrina-Parana, Brazil}
\author{Luciano Vanzo\femail{vanzo} and
Sergio Zerbini\femail{zerbini}}
\Idinfn


\babs
The thermodinamic properties of the (2+1)-dimensional non-rotating black hole 
of 
Ba\~nados, Teitelboim and Zanelli are discussed. The first quantum correction 
to the Bekenstein-Hawking entropy is evaluated within the on-shell Euclidean 
formalism, making use of the related Chern-Simons representation of 
the 3-dimensional gravity. Horizon 
and ultraviolet divergences  in the quantum correction are dealt 
with a renormalization of the Newton constant. It is argued that the quantum 
correction due to the gravitational field  shrinks the effective radius of a
hole and becomes more and more important as soon as the evaporation process 
goes on, while the area law is not violated.

\eabs

\pacs{04.60.Kz,\,\,\,\,\,04.70.Dy,\,\,\,\,\,97.60.Lf}

\s{Introduction}

It is well known that we do not have yet at disposal a consistent and
complete 4-dimensional quantum gravity,
but nevertheless a large number of interesting issues have been
investigated, mainly within the semiclassical approximation. One of
the most important issue is related to the black hole physics and
deals with the origin of the entropy, its quantum corrections,
the information loss paradox, the validity of the area law  (see, for example
Ref. \cite{beke94u-15}). However, it is well known that in (3+1)-dimensions, 
the black hole quantum physics  needs several approximations. 

Recently 3-dimensional gravity has been studied in detail. Despite the
simplicity of the 3-dimensional case (no propagating gravitons), it is a
common believe that it deserves attention  as useful laboratory. In fact with
surprise a black hole solution has been found by Ba\~nados, Teitelboim and
Zanelli \cite{bana92-69-1849}, the so called BTZ black hole. In particular, a
simple geometrical structure of a black hole allows exact computations since
its Euclidean counterpart is locally isomorphic to the constant curvature
3-dimensional hyperbolic space $H^3$.

In this paper we shall compute the first quantum correction to the
semiclassical Bekenstein-Hawking entropy for the BTZ black hole due to 
the one-loop gravitational fluctuations, in an attempt to elucidate 
the statistical origin of the black hole entropy \cite{beke73-7-2333,hawk75-43-199,gibb77-15-2752,frol94-74-3319} and to 
explore the possible relevance  of the quantum fluctuations during the 
late stage of the black hole evaporation process.

With regard to this issues, we recall that a lot of papers have been 
appeared where the quantum entropy of matter fields, propagating in a 
black hole background, has been evaluated by means of several different 
techniques 
(see, for example, Refs. 
\cite{thoo85-256-727,dowk,ghos,suss,solo,furs,cogn95-52-4548,cogn95-12-1927,barv95-51-1741,frol,byts96-458-267} 
and reference therein).
We would like to stress here that we shall compute the one-loop contribution 
due to the
quantization of the gravitational field itself. The tree level 
approximation to the partition function as been discussed at length in 
\cite{mann94}, using the Brown and York approach to quasi-local 
thermodynamic for asymptotically anti-de Sitter black holes. It is 
found that in $(2+1)$ dimensions, there is a 
thermodynamically stable black hole solution, and no negative heat capacity 
instantons. Thus one expects that first quantum 
corrections be well defined.

As far as the computation of these corrections is concerned, some work has
 been done in 
\cite{carl95-51-622,ghos97-56-3568} and a motivation 
of our paper is to present a detailed and possibly complete discussion on this 
point.

The first quantum correction of the BTZ black hole will be
evaluated making use of the  related Chern-Simons representation of 
the 3-dimensional gravity \cite{witt88-311-46,carl93-10-207}. 
It should be stressed that within 
this approach a preliminary 
statistical mechanics explanation of the Bekenstein-Hawking 
entropy, counting boundary states at the horizon, has been given in Ref. 
\cite{carl95-51-632}. 

The contents of the paper are the following. In Sect. 2 we briefly review a
geometry of the Eclidean BTZ black hole. In Sect. 3 we present a derivation of
the Selberg trace formula, starting from an elementary derivation of the heat-
kernel 
trace  related to the Laplace operator,
necessary for our regularization. In Sect. 4 a computation of
the first quantum correction to the entropy is outlined. The paper ends with
some concluding remarks in Sect. 5. In the Appendix some 
explicit computations are included.

\s{The Euclidean BTZ Black Hole }

Following \cite{carl95-51-622} we summarize here geometrical aspects of
the non-rotating BTZ black hole \cite{bana92-69-1849} which are relevant for
our discussion. In the  coordinates $(t,r,\phi)$, the static
Lorentzian metric reads (8G=1 is assumed for the moment, thus the mass is dimensionless)
\beq
ds_L^2=-\at \frac{r^2}{\si^2}-M \ct dt^2+\at \frac{r^2}{\si^2}-M \ct^{-1}
dr^2+r^2 d\phi^2
\:,\label{lm}\eeq
where $M$ is the standard ADM mass and $\si$ is a dimensional constant. A
direct calculation shows that the above metric is a solution of the
3-dimensional vacuum Einstein equation with negative cosmological constant,
i.e.
\beq
R_{\mu \nu}=2\La g_{\mu \nu}\,, \hs R=6\La=-\frac{6}{\si^2}
\:.\label{ee}\eeq
Thus, the sectional curvature $k$ is constant and negative, namely
$k=\La=-1/\si^2$. The metric (\ref{lm}) has an horizon radius given by
\beq
r_+=\sqrt M \si
\:,\label{hr}\eeq
and  it describes a space-time locally isometric to the anti-de Sitter space.

The Euclidean section is obtained by the Wick rotation $t \to i \tau$  and
reads
\beq
ds^2=\at \frac{r^2}{\si^2}-M \ct d\tau^2+\at \frac{r^2}{\si^2}-M \ct^{-1}
dr^2+r^2 d\phi^2
\:.\label{em}\eeq
Changing the coordinates $(\tau,r,\phi) \to (y, x_1,x_2)$ by means of
\beq
y &=&\frac{r_+}{r} e^{\fr{r_+}{\si}\phi}\:, \nn \\
x_1+ix_2&=& \frac{1}{r} \sqrt{ r^2-r_+^2} \exp\left(i\fr{r_+}{\si^2}\tau
+\fr{r_+}{\si}\phi\right) \:,
\label{cco}\eeq
the metric becomes the one of upper-half space representation of
$H^3$, i.e.
\beq
ds^2=\frac{\si^2}{y^2}\at d^2y+dx_1^2+dx_2^2 \ct
\:.\label{hm}
\eeq
As a consequence, the metric Eq.~(\ref{em})
describes a manifold homeomorphic to the hyperbolic space $H^3$.

It is known that the group of isometries of $H^3$ is  $SL(2,\C)$. We shall
consider a discrete subgroup $\Ga\subset  PSL(2,\C)\equiv SL(2,\C)/\{\pm Id\}$
($Id$ is the identity element), which acts discontinuously at the point $z$
belonging to the extended complex plane $\C\bigcup\{\infty\}$. We recall that
a transformation $\ga\neq Id$, $\ga\in\Ga$, with
\beq
\ga z=\frac{az+b}{cz+d}\,,\hs ad-bc=1\,, \hs a,b,c,d\in\C,
\eeq
is called elliptic if $(\Tr\ga)^2=(a+d)^2$ satisfies $0\leq(\Tr\ga)^2<4$,
hyperbolic if $(\Tr\ga)^2>4$, parabolic if $(\Tr\ga)^2=4$ and loxodromic
if $(\Tr\ga)^2\in\C\backslash\aq0,4\cq$. The element $\ga\in SL(2,\C)$ acts on $z=(y,w)\in H^3$,
$w=x_1+ix_2$ by means of the following
linear-fractional transformation:
\beq
\ga z=\at\frac{y}{|cw+d|^2+|c|^2y^2}\,,\frac{(aw+b)(\bar c \bar
w+\bar d)+a \bar c y^2}{|cw+d|^2+|c|^2y^2} \ct
\label{SS3}
\:.\eeq
The periodicity of the angular coordinate $\phi$ allows to
describe the BTZ black hole manifold as the quotient
${\cal H}^3\equiv H^3/\Ga$, $\Ga$ being a discrete group of isometry
possessing a primitive element $\ga_h \in \Ga$ defined by the identification
\beq
\ga_h(y,w)=(e^{\fr{2 \pi r_+}{\si}} y,e^{\fr{2 \pi r_+}{\si}}w) \sim (y,w)
\:.
\label{yid}
\eeq
According to the Eq.~(\ref{SS3}) this corresponds to the matrix
\beq
\hs \hs  \ga_h= \at
\begin{array}{cc}
e^{\fr{r_+}{\si}} & 0 \\
0 & e^{-\fr{r_+}{\si}}
\end{array}
\ct
\:,
\label{p}\eeq
namely to an hyperbolic element ($\tr \ga_h >2$) consisting in a pure
dilatation. Furthermore, since in the Euclidean  coordinates $\tau$
becomes an angular type variable with period $\be$, one is leads also to the
identification
\beq
\ga_e(y,w)=(y,e^{\fr{i\be r_+}{\si^2}}w)\sim (y,w)
\:.\label{eid}\eeq
This identification is generated by an elliptic element in the group $\Ga$
\beq
\hs \hs \ga_e= \at
\begin{array}{cc}
e^{i\be \fr{r_+}{\si^2}} & 0 \\
0 & e^{-i\be \fr{r_+}{\si^2}}
\end{array}
\ct
\:,\label{p1}\eeq
as soon as $\tr \ga_e <2$, and a conical singularity will be present. If
\beq
\be \frac{r_+}{\si^2}=2 \pi
\:,\label{h}\eeq
then $\ga_e=Id$ and the conical singularity is absent. As a result the period 
is determined to be
\beq
\be_H=2 \pi \frac{\si^2}{r_+}
\:,\label{h1}\eeq
and this is interpreted as the inverse of the Hawking temperature
\cite{gibb77-15-2752}. Therefore the on-shell BTZ black hole can be regarded as a 
strictly hyperbolic non-compact manifold ${\cal H}^3$.  The mass as a 
function of the black hole temperature $T=\be_H^{-1}$ reads
\beq
M=4\pi^2\si^2T^2,
\eeq
which shows the stability condition $\partial M/\partial T>0$ is 
fulfilled. 
 The tree-level Bekenstein-Hawking entropy $S_H$ may be
simply obtained making use of the relation
\beq
\beta_H=\frac{\partial S_H}{\partial M}
\:.\label{bhr}
\eeq
Thus one has
\beq
S_H=4 \pi r_+=2A
\:,\label{bh}\eeq
which is the well known "area law" for the black hole entropy. Note that 
$A=2\pi r_+$ is the perimeter of the horizon. If we choose $G=1$ instead 
of $8G=1$, the entropy becomes $A/4$, as it is more familiar 
to a black hole physicists. 

Another important 
thermodynamics input is the off-shell Euclidean action of a black 
hole, namely the action evaluated at $\be\neq\be_H$ (c.~f. 
\cite{mann94} for the quasi-local formalism of thermodynamics)
\beq
I=-\frac{1}{2\pi}\int_{\ca 
M}({\cal R}-2\La)\sqrt{g}\,d^3x-\frac{1}{\pi}\int_{\partial\ca 
M}{\cal K}\sqrt{h}\,d^2x.
\eeq
The boundary $\partial\ca M=S^1\otimes S^1$ (a torus) is identified with 
period $\be$ (the first circle) at some fixed radius $r=R$ (the second 
circle), which will 
be taken to infinity at the end, and ${\cal K}$ is the trace of the extrinsic 
curvature of the boundary. The Euclidean action (2.18) is a 
divergent function of the boundary location, and therefore it is 
necessary to subtract  the action of a chosen 
background \cite{hawk96-13-1487} from it. This will be the zero mass solution, 
i.e. the $M=0$ line element
\beq
ds^2_0=\frac{r^2}{\si^2}d\tau^2+\frac{\si^2}{r^2}dr^2+r^2d\phi^2,
\label{z0}
\eeq
which corresponds 
also to the zero temperature state; all quantities 
referring to this reference background have a subscript "$0$". 
As mentioned above, at $\be\neq\be_H$ there will be a conical singularity 
whose 
contribution to the action, as is well known, is given by the Gauss-Bonnet 
theorem for a disk \cite{teit95-51-4315}, and is
\beq
\frac{1}{2\pi}\int_{\ca M}{\cal R}\sqrt{g}\,d^3x=\frac{2A}{\be_H}(\be_H-\be).
\eeq
Contribution related to a background is vanishes, since $A_0=0$. The 
difference  of the actions can be computed by matching the 
coordinate of the boundary location in the background, $r=R_0$, to the 
coordinate of the boundary location in the black hole metric, $r=R$, 
so that the two metrics asymptotically agree. Finally, the surface 
contribution is seen to vanish. Therefore, the off-shell 
Euclidean action becomes
\beq
I=-\frac{2A}{\be_H}(\be_H-\be)-\frac{r_+^2\be}{\si^2}=M\be-2A,
\eeq
where $(M,\be)$ are now independent thermodynamics variables and 
$r_+=\si\sqrt{M}$. On-shell we have $\be=\be_H$ and $I=-2\pi r_+$. If 
one identifies $I$ with $-\ln Z$ \cite{gibb77-15-2752}, the partition 
function of the black hole, then the mean energy in the canonical ensemble 
will be
\beq
<E>=-\partial_{\be}\ln Z=M,
\eeq
as it was to be expected, and the entropy will be again $S=2A=4\pi 
r_+=4\pi\si\sqrt{M}$. Because $S\sim\sqrt{M}$, the partition function  
as a "sum over states" in semiclassical quantum $2+1$-gravity will 
converge, and the canonical ensemble for a black hole in equilibrium 
with thermal radiation will lead to a stable thermodynamics.

We conclude this section with a comment on the global geometry of the 
ground state.
Looking at Eq.~(\ref{z0}), it is clear that $\tau$ can be identified to 
any period $\be$ (in particular $\be=\ii$), and that $\phi$ has the period 
$2\pi$. Changing the coordinates as  $r=\si^2/y$, $\tau=x_1$ and 
$\phi=x_2/\si$, one gets  the metric of
hyperbolic space
\beq
ds^2_0=\si^2\,\frac{dx_1^2+dx_2^2+dy^2}{y^2},
\eeq
and the identification 
$\ga_p(w,y)=(w+\be +i2\pi\si\,y) \simeq (w,y)$. This identification is generated by  
elements 
of $\Ga$ of the form   
\beq
 \hs \ga_{p_1}= \at
\begin{array}{cc}
1 & \be \\
0 & 1
\end{array}
\ct \,\,,
\hs \ga_{p_2}= \at
\begin{array}{cc}
1 & 2\pi i \si \\
0 & 1
\end{array}
\ct
\:,\label{par}\eeq
which are parabolic. Thus, our reference manifold can be regarded as 
the quotient ${\cal H}_0=H^3/\Ga_0 $, where a subgroup $ \Ga_0$ has  
primitive parabolic elements $\ga_{p_1}$ and  $\ga_{p_2}$. 

We  note that for negative mass, one gets 
solutions with naked conical singularity \cite{dese84-152-220} unless 
one arrives at $M=-1$, namely $H^3$, the Euclidean counterpart of the 
3-dimensional anti-de Sitter space-time. This solution  is a 
permissible solution and can regarded as a "bound 
state" \cite{bana92-69-1849}.   

\s{ Trace Formula and the Spectral Zeta Function}

In this section we investigate the spectral properties associated with the 
Laplace type 
operator acting in a non-compact hyperbolic 
manifold ${\cal H}^3$. To go further, it is convenient to introduce spherical hyperbolic coordinates
\beq
y=\cos \theta\,, \hs
w=x_1+ix_2=\rho \sin \theta e^{i\varphi}
\:.\label{sph}\eeq
It is easy to show that the fundamental domain of ${\cal H}^3$ is non-compact
and it can be given as follows \cite{carl95-51-622}
\beq
 F=\ag 1 \leq \rho \leq N, 0 \leq \theta <\pi/2, 0 < \varphi <
2\pi \cg
\:,\label{fd}\eeq
where $\ln N=2 \pi r_+/\si$. Note that $z'=\ga_h z=N z$ and the corresponding
transformation law for a scalar field $\Phi$ reads $\Phi(\ga z)=\chi \Phi(z),
\ga \in \Ga$, where $\chi$ is a finite-dimensional unitary representation
(a character) of $\Ga$.

Let us consider an arbitrary integral operator which is given by a kernel
$k(z,z')$. The operator is invariant (i.e. the operator commutes with
all operators of the quasi-regular representation of the group $PSL(2,\C)$
in the space $C_0^{\infty}(H^3)$) if its kernel satisfy the condition
$k(\ga z,\ga z')=k(z,z')$ for any $z,z'\in H^3$. Thus, the kernel of the invariant
operator, for example the Laplace operator, is a function of the geodesic
distance between $z$ and $z'$, namely
\beq
d(z,z')=\cosh^{-1} \aq 1+\frac{(y-y')^2+(x_1-x_1')^2+(x_2-x_2')^2}{2yy'}\cq
\:.\label{gd}\eeq
Then the geodesic length between the point $z$ and $z'=\ga_hz$ is
\beq
l_0=\inf d(z,\ga_h z)=\ln N=2 \pi\frac{r_+}{\si}
\:.\label{gd1}\eeq
It is convenient to replace such a distance with the fundamental
invariant of a pair of points
\beq
u(z,z')=\frac{1}{2}\at \cosh d(z,z')-1 \ct \,,\ \hs u(z,z)=0
\:,\label{u}\eeq
and therefore $k(z,z')=k(u(z,z'))$. Finally for the sake of simplicity we put
$\si=1$ thus $|k|=1/\si^2=1$ and all the quantities are dimensionless (the
physical dimensions can be restored by dimensional analysis at the end of the
calculations).

\ss{The Heat Kernel Trace Formula}

Let us start with the heat kernel  of the
Laplace operator  acting in $H^3$. We shall use the method of images. The heat-kernel reads
(see, for example, \cite{camp90-196-1,byts96-266-1})
\beq
K^{H^3}_t(z,z')=\frac{\exp\left(-t-\fr{d^2(z,z')}{4t}\right)}{(4\pi t)^{\fr{3}{2}}}
\frac{d(z,z')}{\sinh d(z,z')}
\:.\label{hhk}\eeq
With regard to the heat kernel on $ {\cal H}^3$, the method of images gives
\beq
K_t(z,z')= \sum_{n} \chi^n K^{H^3}_t(z,\ga_h^n z')= K^{H^3}_t(z,z')+
\sum_{n \neq 0} \chi^n K^{H^3}_t(z,\ga_h^n z')
\chi^n
\:,\label{im}\eeq
where the separation between the identity and the non-trivial periodic geodesic
contribution has been done. In our case, the volume $V(F_3)$ of the fundamendal
domain $F_3$ is divergent and we must introduce a regularization. The simplest
one is to limit the integration in variable $\theta$ between
$0<\theta<\pi/2-\ep$, with $\ep$ suitable. Thus we have
\beq
V_\ep(F)=\int_1^N \frac{d\rho}{\rho}\int_0^{2\pi}
d\varphi \int_0^{\pi/2-\ep}\frac{\sin \theta}{(\cos
\theta)^3}d\theta=2\pi^2 r_+(\cot \ep)^2 =2  \pi^2 r_+ \at 
\frac{1}{\ep^2}-\frac{2}{3}+{\cal O}(\ep)\ct
\:.\label{fdv}\eeq
We may determine $\ep$ choosing 
\beq
\frac{1}{\ep^2}=\frac{R^2}{r_+^2}-\frac{1}{3}
\:.\label{baa}\eeq
Thus
\beq
V_R(F)=2\pi^2 \frac{R^2}{r_+}-2\pi^2 r_+=\int_0^{\be_H} d\tau 
\int_0^{2\pi} d\phi 
\int_{r_+}^R r dr
\:,\label{fdp1}\eeq
where the cuttof parameter $R$ has been introduced (see Sect. 2 for notation). 
The integration over the regularized (fundamental) domain of the diagonal part
leads to
\beq
\mbox{Tr}e^{-t\lap_0}(R)&\equiv&\Tr K_t(R)\,=\,V_R(F)\frac{e^{-t}}{(4\pi
t)^{\fr{3}{2}}}\nn \\
&+&2 \pi l_0  e^{-t} \sum_{n \neq 0}  \frac{\chi^n}{(4\pi t)^{\fr{3}{2}}}
\int_0^{\pi/2} 
\frac{(\sin \theta) d(z,\ga_h^n z)}{(\cos
\theta)^3\sinh[d(z,\ga_h^n z)]}  e^{-\fr{d^2(z,\ga_h^n z)}{4t}}d\theta
\:,\label{bn1}\eeq
where $\lap_0$ is a scalar Laplacian, $d(z,\ga_h^n z)=\cosh^{-1} 
(1+b_n^2\cos^{-2} \theta)$. The integral over
$\theta$ can be performed by means of change of the integration variable
$\theta \to u$ given by $2\sqrt {u t}=\cosh^{-1} (1+b_n^2\cos^{-2} \theta)$.
As a consequence, the resulting integral becomes elementary, i.e.
\beq
\Tr K_t(R)= V_R(F)\frac{e^{-t}}{(4\pi
t)^{\fr{3}{2}}}
+4 \pi l_0 \sum_{n=1}^\ii   \frac{\chi^n e^{-t}}{b_n^2(4\pi t)^{\fr{3}{2}}}
\int_{\frac{n^2 l_0^2}{4t}}^{\ii}e^{-t u}du
\:,\label{bn6}\eeq
since $\cosh^{-1} (1+b_n^2)=n l_0$. As a result one obtains
\beq
\Tr K_t(R)=V_R(F)\frac{e^{-t}}{(4\pi
t)^{\fr{3}{2}}}+\frac{l_0}{2} \sum_{n=1}^\ii  \frac{\chi^n}{(\sinh \fr{n
l_0}{2})^2}\frac{e^{-t-\fr{l_0^2n^2}{4t}}}{(4\pi t)^{\fr{1}{2}}}
\:.\label{hk}\eeq
Recently, the above heat-kernel trace has also been computed in 
\cite{mann}.

\ss{The Explicit Form of the Zeta Function}

In the previous subsection we have derived the heat kernel trace formula.
For our purpose it is important that 
Eq. (\ref{hk}) looks (formally) as the Selberg trace formula associated
with Laplace operator acting in a compact space ${\cal H}^3$ (a group $\Ga$
is co-compact). This statement is formal enough, nevertheless let us verify it
withstanding a common style of presentation. 

First of all we may consider a given (regularized) compact Riemannian 
manifold as conformally equivalent to one of constant scalar curvature. This 
is known as the Yamabe problem \cite{yama60-12-21}. This problem has been 
solved for the case of non-positive scalar curvature in Ref. 
\cite{trud68-3-265}. Furthermore, let $\{\la_j\}_{j=0}^{\infty}$ denote the 
non-zero isolated eigenvalues (appearing the same number of times as its 
multiplicity) of positive self-adjoint Laplace operator. Let us introduce a 
suitable analytic function $h(r)$, where $r^2_j=\la_j-1$. It can be shown that
$ h(r)$ is related to the quantity $k(u( z,\ga z))$ by means of the Selberg 
transform (see for example \cite{eliz94b,byts96-266-1} and references 
therein). Let $\hat{h}(p)$ being the Fourier transform of $h(r)$,
\beq
\hat{h}(p)=\frac{1}{2\pi}\int_{-\ii}^\ii e^{-irp} h(r) dr
\:.\eeq
For the derivation of the Selberg trace formula, one has to consider the
contributions coming from the identity element in $\Ga$ and all $\ga$-type
conjugacy classes (the metod of images), namely
\beq
\Tr h(\lap_0)=\sum_j h(\la_j)={\cal C}(I)+{\cal C}(H)=
V(F_3)k(0)
+\sum_{\{\ga\}} \chi(\ga)\int_{F_3} k(u(z,\ga z)) d\mu_3
\:.\label{stf}\eeq
The first term in the r.h.s. of Eq.~(\ref{stf}) ${\cal C}(I)$ is the
contribution of the identity element, while $V(F_3)$ is the (finite) volume
of the fundamental domain with respect to the Riemannian measure $d\mu_3=
dx_1dx_2dy y^{-3}$. Formally for the non-compact manifold ${\cal H}^3$, 
whose fundamental domain is given by Eq.~(\ref{fd}), one may put
$V(F_3)\sim V_R(F)$, where $V_R (F)$ is given by Eq.~(3.10).

Let us consider now the hyperbolic (a topologically non-trivial) contribution
and show that it is finite. First it reduces to
\beq
{\cal C}(H)=\sum_{\{\ga\}} \chi(\ga)\int_{F_3} k(u(z,\ga z)) d\mu_3=
\sum_{n\neq 0}\chi^n \int_{F_3} k(u(z,\ga_h^n z))d\mu_3
\:.\label{top}\eeq
Noting that $\chi^n=\chi^{-n}$ and
\beq
u(z,\ga_h^nz)=\frac{1}{2}\at \cosh d(z,\ga_h^nz)-1 \ct=b_n^2 (1+\tan^2
\theta)
\:,\label{van}\eeq
with $b_n^2= \sinh^2 (\fr{nl_0}{2})$, one has
\beq
{\cal C}(H) = 4\pi l_0 \sum_{n=1}^\ii \chi^n  \int_0^{\pi/2} \frac{\sin 
\theta}{(\cos\theta)^3}k \at b_n^2 (1+\tan^2 \theta) \ct d\theta
= 2  \sum_{n=1}^\ii \frac{\chi^n}{b_n^2} \int_{b_n^2}^\ii 
k(x) dx
\:.\label{t}\eeq
Recalling the Selberg transform in the 3-dimensional case
\cite{eliz94b,byts96-266-1} one gets
\beq
k(0)&=&\int_0^\ii\frac{r^2}{2\pi^2}\:h(r)\:dr\:, \nn \\
\int_{b_n^2}^\ii k(x) dx &=& \frac{1}{4\pi}\hat{h}(n 2 r_+)
\:.\label{st}\eeq
Thus the final trace formula reads
\beq
\Tr h(\lap_0)(R) =
V_R(F) \int_0^\ii\frac{r^2}{2\pi^2}\:h(r)\:dr
+ l_0 \sum_{n=1}^\ii \chi^n \frac{\hat{h}(n 2 r_+)}{(\sinh \fr{nl_0}{2})^2}
\:.\label{stf2}\eeq
The trace formula (\ref{stf2}) is valid for a large class of $h(r)$ function.
In particular, choosing $h(r)=e^{-t(r^2+1)}$ in Eq.~(\ref{stf2}), one 
obtains the result of Eq.~(\ref{hk}).

Finally the related zeta function can be calculated by means of the Mellin
transform
\beq
\zeta(s|\lap_0)(R)=\frac{1}{\Ga(s)}\int_0^\ii t^{s-1}\Tr K_t(R) dt
\:,\label{mt}\eeq
valid for $\Re s> 3/2$.
A direct computation gives the analytic continuation of the zeta function
in neighborhood of point $s =0$, i.e.
\beq
\zeta(s|\lap_0)(R)=V_R(F)
\frac{\Ga(s-\fr{3}{2})}{(4\pi)^{\fr{3}{2}}\Ga(s)}+\frac{l_0}{\Ga(s)\Ga(1-s)}
\int_0^\ii \at 2t+t^2 \ct^{-s}\Psi(2+t) dt
\:,\label{z111}\eeq
where the function
\beq
\Psi(s)=\sum_{n=1}^\ii  \frac{\chi^n}{(\sinh \fr{n
l_0}{2})^2}e^{-(s-1)l_0n}
\:,\label{por}\eeq
has been introduced.

For transverse 1-forms, there exists a similar trace formula, (see for 
example, \cite{byts97u-398}) and we quote here only the results: there is 
no gap in the spectrum of Laplace operator $\lap^\perp_1$; the 
Plancherel measure is $(r^2+1)/(2\pi^2)$ and the heat-kernel trace formula 
reads
\beq
\Tr e^{-t\lap^\perp_1}(R) =
Q(t) \at V_R(F)+ 
+ 2 l_0 (4\pi t)\sum_{n=1}^\ii \chi^n \frac{e^{-\fr{n^2l_0^2}{4t}}}{(\sinh \fr{nl_0}{2})^2}
\ct
\:,\label{pert}\eeq
with
\beq
Q(t)=\frac{1}{4(\pi 
t)^{\fr{3}{2}}}+\frac{1}{2\pi^{\fr{3}{2}}t^{\fr{1}{2}}}
\:.\label{q}\eeq

\section{The First Quantum Correction to the Entropy of the BTZ Black Hole}

The first quantum correction to the Bekenstein-Hawking entropy may be computed
within the Euclidean semiclassical approximation  \cite{gibb77-15-2752} and
we shall follow this approach in this section. We have to mention that a more
sophisticate approach has been proposed in Ref. \cite{brow93-47-1420}, where
the canonical and microcanonical partition function of the black hole in a 
cavity with suitable boundary conditions has been investigated. This 
approach has the merit of a more direct physical understanding, and 
has been applied to anti-de Sitter black holes in \cite{mann94}.

Within the Euclidean approach, making use of the Chern-Simons 
representation of the 3-dimensional gravity 
\cite{witt88-311-46,carl93-10-207}, the one-loop approximation gives
\beq
\ln Z^{(1)}=\ln (T^{1/2})-I
\:,\label{part0}\eeq
where one is dealing with a compact 3-manifold $M$ and the quantum prefactor 
$T$ is the Ray-Singer torsion associated with $M$ (see a more precise 
definition below). 
In our case, we assume the quantum prefactor to be the same, but
\beq
\ln Z^{(1)}=\frac{1}{2}\ln T-(I_{BTZ}-I_{0})-(B_{BTZ}-B_{0})
\equiv \frac{1}{2}\ln T-I_P
\:,\label{vanzo}\eeq
in which  $B_{BTZ}$ is the usual
boundary term which depends on the extrinsic curvature at large spatial
distance. The total classical action is divergent; the geometry is 
non-compact and we have introduced the "reference" background ${\cal H}_0^3$ 
at the tree level \cite{hawk96-13-1487} and the related  volume cutoff 
$R$ and $R_0$.
With this proposal, in Eq. (4.2) the two boundary terms of the classical 
contribution cancel for large $R$ and the difference of
the on-shell Euclidean classical actions gives rise to (see Sect. 2), 
\beq
I_P=I_{BTZ}-I_{0}=-\frac{2}{\pi} \at V(R)-V_0(R) \ct \to -2 \pi r_+=
-\ln Z^{(0)}
\:.\label{bh1}\eeq

Restoring the correct physical dimension in Eq. (4.3), it is easy to show that 
the on-shell tree-level partition function $ Z^{(0)}$, Eq.~(\ref{bh1}), 
becomes
\beq
\ln Z^{(0)}=\frac{4 \pi^2 r_+}{16 \pi G}
\:,\label{zoo}\eeq
and this leads to
the semiclassical Bekenstein-Hawking entropy
\beq
S^{(0)}=S_H=\at r_+\frac{\partial}{\partial r_+}+1 \ct \ln Z^{(0)}=
\frac{1}{4}\frac{2 \pi r_+}{G}
\:.\label{bh11}\eeq

So far we have neglected quantum fluctuations. The first on-shell quantum correction 
of the gravitational quantum fluctuations is given by  the square root of the
Ray-Singer torsion of the manifold ${\cal H}^3$. 
For a compact hyperbolic manifold the Ray-Singer
torsion is the ratio between functional determinants of Laplace operators
$\lap_k$ acting on $k$-forms on ${\cal H}^3$ (see for example  
\cite{witt88-311-46,carl93-10-207,byts97u-398}, i.e.
\beq
T=\frac{\det \lap_0}{(\det \lap^\perp_1)^{1/2}}
\:.\label{rst}\eeq
However, in our case a manifold is non-compact and a 
volume regularization previously introduced will be used. 
Thus, we have
\beq
\ln Z^{(1)}=\frac{1}{2}\ln \det \lap_0-\frac{1}{4}\ln \det \lap^\perp_1-I_P  
\:.\label{ziii}\eeq
For the tree level term it is necessary to introduce the bare 
quantity $G_B$, since the quantum correction are plagued by the 
ultraviolet divergences and a renormalization procedure must be used.  The 
functional determinants are 
then calculated by means of a regularization. We shall use  
the proper-time regularization (zeta-function regularization gives the 
same finite part), in order to deal explicitly with the 
ultraviolet divergences. 

In the case of $0$-forms, one can compute a functional determinant by means of
Eq.~(\ref{hk}). Thus, we have
\beq
\ln \det \lap_0 &=& -\int_\ep^\ii t^{-1} \Tr e^{-t \lap_0} dt=-\frac{V_R}{(4 \pi)^{3/2}}
\Ga(-\fr{3}{2},\ep)+
\sum_{n=1}^\ii  \frac{\chi^n}{n(\sinh \fr{n l_0}{2})^2} e^{-l_0n} \nn  \\ 
&=&-\frac{V_R}{(4 \pi)^{3/2}}\Ga(-\fr{3}{2},\ep)+\ln {\cal Z}_0(2)
\:,\label{z1}\eeq
where $\Ga(-\fr{3}{2},\ep)$ is the incomplete Gamma function, which 
has two divergent terms as $\ep \to 0$, namely
\beq
\Ga(-\fr{3}{2},\ep)=\Ga(-\fr{3}{2})-\frac{1}{4(\pi 
\ep)^{\fr{3}{2}}}+\frac{1}{(4\pi)^{\fr{3}{2}}\ep^{\fr{1}{2}}}+{\cal O}
(\ep^{\fr{1}{2}})
\:,\label{ig}\eeq
and
\beq
\ln{\cal Z}_0(2)=\sum_{k=1}^\ii k \ln \at 1-\chi
e^{-2(k+1)\fr{r_+}{\si}} \ct
\:.\label{buu}\eeq

In a similar way, using Eq.~(\ref{pert}), one has
\beq
\ln \det \lap^\perp_1=-\int_\ep^\ii t^{-1} \Tr e^{-t 
\lap^\perp_1} dt=-\frac{V_R}{8(4\pi 
\ep)^{\fr{3}{2}}}+\frac{V_R}{2(4\pi)^{\fr{3}{2}}\ep^{\fr{1}{2}}} 
-\ln{\cal Z}_1(1)
\:,\label{z2}\eeq
with
\beq
\ln{\cal Z}_1(1)=\sum_{n=1}^\ii \frac{\chi^n}{\left(\sinh (n \fr{r_+}{\si})
\right)^2} \at
\frac{1}{n}+8 n (\fr{r_+}{\si})^2 \ct
\:.\label{buu1}\eeq
As a result
\beq
\ln Z^{(1)}=\frac{ 4\pi^2 r_+}{16 \pi G}+g(r_+)-F_\ep
\:,\label{bh12}\eeq
where
\beq
g(r_+)=\frac{1}{2}\ln{\cal Z}_0(2)+\frac{1}{4}\ln{\cal Z}_1(1)
\:,\label{ggg}\eeq
and
\beq
F_\ep=V_R \at \frac{1}{2(4 \pi)^{3/2}}\Ga(-\fr{3}{2},\ep)+\frac{1}{32(4\pi 
\ep)^{\fr{3}{2}}}-\frac{1}{8(4\pi)^{\fr{3}{2}}\ep^{\fr{1}{2}}} \ct
\:.\label{fe}\eeq
If we define the renormalized quantity
\beq
\frac{1}{16 \pi G_r}=\frac{1}{16 \pi G}+\frac{F_\ep}{4\pi^2 r_+}
\:,\label{rg}\eeq
we arrive at
\beq
\ln Z^{(1)}=\frac{\pi r_+}{4 G_r}+g(r_+)
\:.\label{bh123}\eeq
This renormalized one-loop effective action may be thought to describe an 
effective classical 
geometry belonging to the same class 
of the non rotating BTZ black hole solution. This stems from  the 
results contained in \cite{henn}, where it has been shown that the 
constraints for pure gravity have an unique solution. As a consequence, one may 
introduce a  new  effective radius by means of 
\beq
\ln Z^{(1)}=\frac{\pi R_+}{4 \pi G_r}
\:,\label{bbb}\eeq
where
\beq
R_+=r_++ \frac{4G_r}{\pi} g(r_+)
\:,\label{r}\eeq
mimicking in this way the back reaction of the quantum gravitational 
fluctuations. As a consequence, the new  entropy  is given 
by an effective Bekenstein-Hawking term, namely 
\beq
S^{(1)}=\frac{1}{4}\frac{ 2\pi R_+}{G_r}
\:.\label{}\eeq

One can evaluate the asymptotics of the quantity $g(r_+) $ for 
$r_+ \to \ii$ and  
$r_+ \to 0$ and then compute the  effective radius. Note that 
$ \ln{\cal Z}_1(1)$ and $\ln{\cal Z}_0(2)$ are esponentially small for 
large  $r_+$. Thus
\beq
R_+ \simeq r_+
\:,\label{bnm}\eeq
and nothing of interesting is present in this limit. 

Making use of the results of 
the Appendix, for small $r_+$, one has
\beq
R_+ &\simeq & r_++\frac{4G_r}{\pi} \aq \frac{\si^2}{16r_+^2} \at 
-\ln \left(\frac{2 r_+}{\si}\right)+  2\ga + 
\Psi(2)-\ze(3) \ct \cp \nn\\
&+& \ap\frac{\si \pi^2}{24 r_+}+\frac{1}{4} \ln \left( \frac{r_+}{\si 
\pi}\right)+ {\cal O}(r_+)\cq
\:.\label{bnm1}\eeq
One can see that for $r_+$ sufficiently small the effective radius 
becomes larger and positive. This means that $R_+$ (as a function of $r_+$) 
reaches a minimum for suitable $r_+^*$, solution of the equation
\beq
\frac{4G_r}{\pi}  g'(r_+^*)=-1
\:.\label{*}\eeq
This result is in qualitative agreement with a very recent computation of the 
off-shell quantum 
correction to the entropy due to a scalar field in the BTZ background 
\cite{mann} 
and all the qualitative considerations contained there are also valid for the 
gravitational case we are dealing with. In particular, it appears that the 
quantum gravitational 
corrections could become more and more important as soon as the evaporation 
process continues and thus they cannot be neglected.

\section{Concluding Remarks}

In this paper the first quantum correction of the BTZ black hole has been
evaluated making use of the  appropriate Chern-Simons representation of 
the 3-dimensional gravity. The quantum prefactor, i.e. the Ray-Singer 
torsion, has been evaluated 
by means of the proper-time  regularization. 

In our computation the one-loop ultraviolet and horizon divergences, 
generally present
in the first quantum correction, have also been found and 
they have been accounted for
by means of the introduction of the standard  one-loop renormalization procedure 
of the Newton constant \cite{suss}. The semiclassical
Bekenstein-Hawking entropy has been also rederived by the improved  
Euclidean method suggested in \cite{hawk96-13-1487}. 

Our result for quantum corrections differs from the one
reported in Ref. \cite{carl95-51-622} and  are consistent
with the detailed computation of the entropy of scalar fields in a BTZ 
classical backgrounds given in Ref. \cite{mann}. With  regard to this, 
horizons divergences 
of the entropy for scalar fields in the same background have  also 
been investigated in \cite{ichi95-447-340}.

Finally, our result, even though obtained in the one-loop 
approximation may be interpreted with a non violation of the area 
law, but with an effective radius which is the classical one for large 
black hole mass, but which shrinks, as soon as the black hole 
evaporation goes on. This seems to suggest the quantum corrections of the 
gravitational field 
become more and more important near the end of the evaporation process. 
As far as this issue is concerned, we observe 
that the final effective geometrical configuration is the reference 
space  ${\cal H}_0^3$, which admits a naked singularity at the origin. 
As a consequence, the quantum correction seems to have a tendency to avoid the 
appearance of 
the naked singularity, in agreement with the "cosmic censorship" 
hypothesis.

\ack{A.A. Bytsenko wishes to thank CNPq and the Department of Physics of
Londrina University for financial support and kind hospitality. The research 
of A.A. Bytsenko ~was supported in part by Russian Foundation for
Fundamental Research grant No.~95-02-03568-a and by Russian Universities grant
No.~95-0-6.4-1.}

\section{Appendix }

In this Appendix we shall investigate the small $t=2r_+/\si$ asymptotics for 
the 
quantity 
$g(t)$, making use of the standard Mellin transform technique 
\cite{eliz95-28-617}. For the sake of simplicity we put $\chi=1$. To begin 
with, 
we observe that $\ln {\cal Z}_0(2)$ may be rewritten as
\beq
\ln{\cal Z}_0(2)=\sum_{n=1}^\ii n \ln \at 1-
e^{-t n} \ct-\H (t)
\:,\label{buuh}\eeq
where $\H (t)$ is the Hardy-Ramanujan modular function, given 
by
\beq
\H (t)=\sum_{n=1}^\ii \ln \at 1-e^{-t n} \ct
\:.\label{hr1}\eeq 
It satisfies the functional equation
\beq
\H (t)=-\frac{\pi^2}{6t}-\frac{1}{2} \ln \at \frac{ t}{2 
\pi}\ct+\frac{t}{24}+\H \left(\frac{4\pi^2}{t}\right)
\:.\label{hr12}\eeq
For the first term, the Mellin transform representation gives
\beq
\sum_{n=1}^\ii n \ln \at 1-
e^{-t n} \ct=-\frac{1}{2\pi i}\int_{\Re z>2} t^{-z} 
\Ga(z)\ze(z+1)\ze(z-1) dz
\:.\label{mtr1}\eeq
Shifting the vertical contour to the left, one has a simple pole at 
$z=2$, a double pole at $z=0$ and simple poles at  $z=-2m$, 
$m=1,2,...$ . The Residue theorem gives for small $t$ 
\beq
\sum_{n=1}^\ii n \ln \at 1-e^{-t n} \ct= -\frac{\ze(3)}{t^2}+\ze(-1) \ln 
t-\ze'(-1)+{\cal O}(t^2)\,.
\eeq
With regard to the quantity $\ln{\cal Z}_1(1)$, the same technique gives
\beq
\ln{\cal Z}_1(1)&=&\sum_{n=1}^\ii \frac{1}{(\sinh n \fr{t}{2})^2} \at
\frac{1}{n}+2 n t^2 \ct= \frac{1}{2\pi i}\int_{\Re z>2} dz t^{-z} 
\Ga(z)\ze(z-1) \at \ze(1+z)+2 t^2\ze(z-1) \ct  \nn \\
&=&\frac{\ze(3)}{t^2}-\ze(-1) \ln 
t+\ze'(-1)+\frac{1}{t^2}\at 
2\ga+\Psi(2)-\ln t \ct +{\cal O}(t^2)
\:.\label{buu2}\eeq

As a result, for small $t$ the asymptotics for the quantity $g(t)$ 
(see Eq.~(\ref{ggg})) reads 
\beq
g(t)=\frac{1}{ 4 t^2} \aq -\ln t+  2\ga + \Psi(2)-\ze(3) \cq
+\frac{\pi^2}{12 t}+\frac{1}{4} \ln \at \frac{t}{2 
\pi}\ct+\frac{t}{48}+{\cal O}(t^2)
\:.\label{asgt}\eeq

\end{document}